\documentclass[aps,prl,twocolumn,showpacs]{revtex4-1}%
\usepackage{amsfonts}
\usepackage{amsmath}
\usepackage{amssymb}
\usepackage{graphicx}%
\providecommand{\U}[1]{\protect\rule{.1in}{.1in}}

\begin{document}
\title{Collective Nuclear Stabilization by Optically Excited Hole in Quantum Dot}
\author{Wen Yang}
\affiliation{Center for Advanced Nanoscience, Department of Physics, University of
California San Diego, La Jolla, California 92093-0319, USA}
\author{L. J. Sham}
\affiliation{Center for Advanced Nanoscience, Department of Physics, University of
California San Diego, La Jolla, California 92093-0319, USA}

\begin{abstract}
We propose that an optically excited heavy hole in a quantum dot can drive the
surrounding nuclear spins into a quiescent collective state, leading to
significantly prolonged coherence time for the electron spin qubit. This
provides a general paradigm to combat decoherence by environmental control
without involving the active qubit in quantum information processing. It also
serves as a unified solution to some open problems brought about by two recent
experiments [X. Xu \textit{et al.}, Nature \textbf{459}, 1105 (2009) and C.
Latta \textit{et al.}, Nature Phys. \textbf{5}, 758 (2009)].

\end{abstract}

\pacs{78.67.Hc, 72.25.-b, 71.70.Jp, 03.67.Lx, 05.70.Ln}
\maketitle

The hole, the removal of an electron from the fully occupied valence band, is
an elementary excitation in semiconductors. Because of the lower orbital
symmetry, the hole in the valence band has properties different from the
electrons in the conduction band. In a quantum dot (QD), the spin of a
localized hole is coupled to the spins of the surrounding atomic nuclei of the
host lattice through\ the long-range dipolar hyperfine interaction, as opposed
to the stronger contact hyperfine interaction between the electron spin and
the nuclear spins~\cite{AbragamBook}. The hole-nuclear coupled system forms a
different paradigm of the central spin model~\cite{SpinBath}, which has been
of interest for a long time in spin resonance spectroscopy, quantum
dissipation, and classical-quantum crossover \cite{Decoherence}. Currently,
the central hole spin in the QD has attracted increasing interest due to its
potential usage as carriers of long-lived quantum information~\cite{HoleQC}
and efforts are being directed towards understanding its interactions with the
environments~\cite{HoleEnvironment1}, especially the nuclear
spins~\cite{HoleEnvironment2}. Recently, a symmetric hysteretic broadening of
the absorption spectrum was observed and attributed to a nuclear polarization
transient\textit{ }induced by an optically excited hole through a
non-collinear term of the dipolar hyperfine interaction
\cite{XuNature09,LaddPRL10}.

Compared with the newly initiated hole spin based quantum technology, the
electron spin based quantum technology has achieved a higher degree of
maturity~\cite{ElectronQC}. One critical obstacle is the short electron spin
coherence time $T_{2}^{\ast}$ due to the randomly fluctuating nuclear field
produced by the \textquotedblleft noisy\textquotedblright\ nuclear
spins~\cite{DecoherenceNuclei}. Prolonging the electron spin coherence time is
of paramount importance in quantum information processing. For this purpose,
two major approaches are under rapid development. One aims at decoupling a
general qubit from the environments by repeated pulsed control of the qubit
\cite{DD}. The other aims at suppressing the fluctuation of the nuclear field,
i.e., stabilizing the nuclear spins, by inducing a steady-state nuclear
polarization~\cite{LossDNP} through the electron-driven
Overhauser~\cite{OverhauserPR53} and reverse
Overhauser~\cite{reverseOverhauser} effects. The latter approach, being
specifically designed for the QD electron spin, has the merit that once the
nuclear spin environment is stabilized, it remains \textquotedblleft
quiet\textquotedblright\ for a long time for electron spin manipulation.
Intensive research~efforts have led to successful nuclear stabilization in QD
ensembles \cite{GreilichScience07} and suppression of the fluctuation of the
nuclear field difference between two neighboring QDs \cite{ReillyScience08}.

Recently, three groups \cite{XuNature09,LattaNaturePhys09,VinkNaturePhys09}
reported significant nuclear stabilization in single QDs, most relevant for
quantum information processing. Vink \textit{et al. }\cite{VinkNaturePhys09}
deduced the stabilization from the observed electron spin resonance locking,
attributed to the electron-driven reverse Overhauser effect
\cite{reverseOverhauser}. Xu \textit{et al.} \cite{XuNature09} and Latta
\textit{et al.} \cite{LattaNaturePhys09} directly observed the stabilization
by optical excitation of trion and blue exciton (both containing a hole spin)
in the Voigt and Faraday geometries, respectively. Xu \textit{et al. }also
observed the resulting prolonged electron spin coherence time. However, two
key observations, the nuclear stabilization observed by Xu \textit{et al.} and
the bidirectional hysteretic locking of the blue exciton absorption peak
observed by Latta \textit{et al.} cannot be explained by existing
theories~\cite{XuNature09,LaddPRL10,OverhauserPR53,reverseOverhauser,LattaNaturePhys09}%
. For a complete understanding of the fundamental physics and future
optimization of the nuclear stabilization, a new theory is needed.

In this Letter, we link the hole spin based quantum technology with the
electron spin based quantum technology by constructing a theory showing that
an optically excited central hole spin can stabilize the nuclear spins and
significantly prolong the coherence time of the electron spin qubit. This
provides a general paradigm to combat decoherence by environmental control
without involving the active qubit. The essential ingredients of this theory
include a finite nuclear Zeeman splitting and the non-collinear dipolar
hyperfine interaction \cite{XuNature09}, through which the fluctuating hole
spin stabilizes the nuclear spins by driving them from the \textquotedblleft
noisy\textquotedblright\ thermal equilibrium state into a \textquotedblleft
quiet\textquotedblright\ collective state, where the strong thermal
fluctuation is largely cancelled by the inter-spin correlation. This enables a
flexible control of the nuclear fluctuation and hence the coherence time of
the electron spin qubit by engineering the hole spin fluctuation spectrum
through the highly developed coherent optical techniques. It also provides a
unified explanation to the two distinct experimental observations: the
appearance of the \textquotedblleft quiet\textquotedblright\ state explains
the observation of Xu \textit{et al.} in the single pump
experiment~\cite{ImamogluNarrowing}, while the switch between the
\textquotedblleft noisy\textquotedblright\ state and the \textquotedblleft
quiet\textquotedblright\ state explains the observation of Latta \textit{et
al. }on the blue exciton \cite{LattaNaturePhys09}.

\begin{figure}[ptb]
\includegraphics[width=\columnwidth]{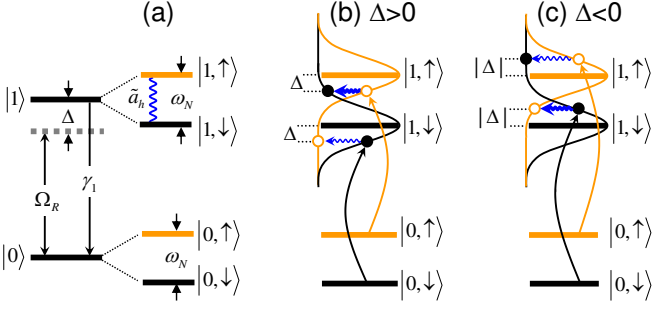}\caption{(color
online). (a)\ Energy levels for the optically excited hole coupled to a
typical nuclear spin-1/2. (b) and (c): Two competing nuclear spin-flip
channels for (b) $\Delta>0$ and (c) $\Delta<0$.}%
\label{G_Setup}%
\end{figure}

The first step of the hole-driven collective nuclear stabilization is a
steady-state nuclear polarization induced by the optically excited hole. The
essential physics of this step is captured by the dynamics of a typical
nuclear spin-1/2 $\hat{\mathbf{I}}$ in a QD coupled to a heavy hole state
$\left\vert 1\right\rangle $ through the non-collinear dipolar hyperfine
interaction $\hat{\sigma}_{11}\tilde{a}_{h}(\hat{I}^{+}+\hat{I}^{-}%
)$~\cite{XuNature09}, where $\hat{\sigma}_{ji}\equiv\left\vert j\right\rangle
\left\langle i\right\vert $, $\hat{I}^{\pm}\equiv\hat{I}^{x}\pm i\hat{I}^{y}$,
and $\tilde{a}_{h}\equiv O(\eta^{2})a_{h}$, with $\eta$ being the heavy-light
hole mixing coefficient. An external magnetic field along the $z$ axis gives
rise to a finite nuclear Zeeman frequency $\omega_{N}$ and a continuous wave
laser couples the ground state $\left\vert 0\right\rangle $ to the excited
hole state $\left\vert 1\right\rangle $ with Rabi frequency $\Omega_{R}$ and
detuning $\Delta$. The hole state dephases with rate $\gamma_{2}$ and decays
back to the ground state with rate $\gamma_{1}$ [Fig.~\ref{G_Setup}(a)]. The
hole dephasing broadens $\left\vert 1,\uparrow\right\rangle $ and $\left\vert
1,\downarrow\right\rangle $ to Lorentzian distribution $L^{(\gamma_{2}%
)}(E)=(\gamma_{2}/\pi)/(E^{2}+\gamma_{2}^{2})$. In the weak pumping limit, two
competing nuclear spin-flip channels $\left\vert 0,\downarrow\right\rangle
\overset{\Omega_{R}}{\rightarrow}\left\vert 1,\downarrow\right\rangle
\overset{\tilde{a}_{h}}{\rightarrow}\left\vert 1,\uparrow\right\rangle $
(down-to-up channel) and $\left\vert 0,\uparrow\right\rangle \overset
{\Omega_{R}}{\rightarrow}\left\vert 1,\uparrow\right\rangle \overset{\tilde
{a}_{h}}{\rightarrow}\left\vert 1,\downarrow\right\rangle $ (up-to-down
channel) are opened up to leading order [Figs. \ref{G_Setup}(b) and
\ref{G_Setup}(c)]. For each channel, the transition rate is proportional to
the square of the coupling strength times the final density of states of each
step determined by the energy mismatch. For the down-to-up channel, the
transition rate $W_{+}\propto\Omega_{R}^{2}L^{(\gamma_{2})}(\Delta
)\times\tilde{a}_{h}^{2}L^{(\gamma_{2})}(\omega_{N}+\Delta)$, where $\Delta$
is the energy mismatch and $L^{(\gamma_{2})}(\Delta)$ is the final density of
states for the first step $\left\vert 0,\downarrow\right\rangle \overset
{\Omega_{R}}{\rightarrow}\left\vert 1,\downarrow\right\rangle $, while
$(\omega_{N}+\Delta)$ and $L^{(\gamma_{2})}(\omega_{N}+\Delta)$ are
corresponding quantities for the second step $\left\vert 1,\downarrow
\right\rangle \overset{\tilde{a}_{h}}{\rightarrow}\left\vert 1,\uparrow
\right\rangle $ [Fig.~\ref{G_Setup}(b) or \ref{G_Setup}(c)]. For the
up-to-down channel, the transition rate $W_{-}\propto\Omega_{R}^{2}%
L^{(\gamma_{2})}(\Delta)\times\tilde{a}_{h}^{2}L^{(\gamma_{2})}(\omega
_{N}-\Delta)$. The competition between the two channels establishes an
intrinsic (i.e., in the absence of other nuclear spin relaxation mechanisms)
steady-state nuclear polarization
\begin{equation}
\langle\hat{I}^{z}\rangle_{0}\propto\frac{W_{+}-W_{-}}{W_{+}+W_{-}}%
=-\frac{2\Delta\omega_{N}}{\Delta^{2}+\gamma_{2}^{2}}+O(\varepsilon^{2})
\label{IZ_INTUITIVE}%
\end{equation}
during a time scale characterized by the inverse of the nuclear polarization
buildup rate $\Gamma_{p}\equiv W_{+}+W_{-}=O(\tilde{a}_{h}^{2}\Omega_{R}^{2}%
)$, where $\varepsilon\equiv\omega_{N}/\gamma_{2}\sim0.1$ for a typical QD
under a magnetic field $B=1\ \mathrm{T}$. The hole-driven intrinsic\textit{
}steady-state polarization $\langle\hat{I}^{z}\rangle_{0}$ shows a sign
dependence on the detuning $\Delta$, which is the key to explain the
observation of Latta \textit{et al.}~on blue exciton excitation
\cite{LattaNaturePhys09}. By contrast, the intrinsic steady-state polarization
$\langle\hat{I}^{z}\rangle_{0}=\langle\hat{S}_{e}^{z}\rangle-\langle\hat
{S}_{e}^{z}\rangle_{\mathrm{eq}}$ ($\langle\hat{I}^{z}\rangle_{0}=\langle
\hat{S}_{e}^{z}\rangle_{\mathrm{eq}}$) due to the electron-driven
Overhauser~\cite{OverhauserPR53} (reverse Overhauser~\cite{reverseOverhauser})
effect is equal to the nonequilibrium (equilibrium) part of the electron spin
polarization and is insensitive to the laser detuning, while the hole-driven
nuclear polarization transient $\langle\hat{I}^{z}\rangle_{\mathrm{transient}%
}\propto\Delta$ \cite{XuNature09} or random shift~\cite{LaddPRL10} produces no
intrinsic steady-state nuclear polarization.

For the hole-driven evolution of a typical nucleus with spin $I\geq$1/2 under
a general pumping intensity, we single out the slow dynamics of the diagonal
part $\hat{P}(t)$ of the reduced density matrix $\hat{\rho}_{N}(t)$ of this
nucleus by adiabatically eliminating the fast motion of the hole and the
off-diagonal part and arrive at%
\begin{equation}
\dot{\hat{P}}=-W_{+}[\hat{I}^{-},\hat{I}^{+}\hat{P}]-W_{-}[\hat{I}^{+},\hat
{I}^{-}\hat{P}], \label{EOM_GENERAL}%
\end{equation}
valid up to $O(\tilde{a}_{h}^{2})$, where the transition rates for the
down-to-up and up-to-down channels, $W_{\pm}=C(\mp\omega_{N})$, are determined
by the steady-state hole fluctuation $C(\omega)=\int_{-\infty}^{\infty
}dt\ e^{i\omega t}\langle\hat{\sigma}_{11}(t)\hat{\sigma}_{11}\rangle$.

\begin{figure}[ptb]
\includegraphics[width=\columnwidth]{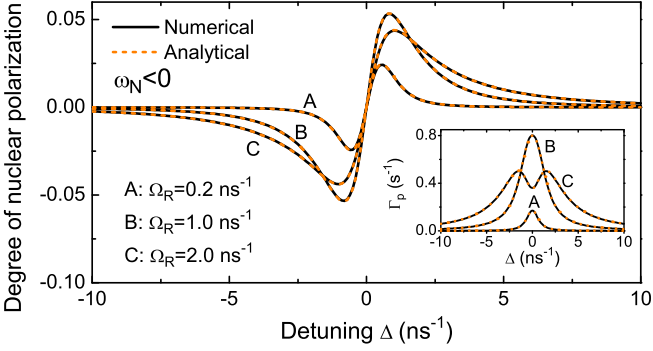}\caption{(color online).
$s^{(\mathrm{ss})}$ and $\Gamma_{p}$ (inset) for a nuclear spin-1/2 obtained
from Eqs. (\ref{EOM_S})-(\ref{IZ_GENERAL}) (dashed lines) compared with the
exact numerical results (solid lines) for $\Gamma_{1}=0.2$ s$^{-1}$,
$\gamma_{1}=\gamma_{2}=1\ \mathrm{ns}^{-1}$, $\omega_{N}=-0.1\ \mathrm{ns}%
^{-1}$, $\tilde{a}_{h}=4\times10^{-5}\ \mathrm{ns}^{-1}$ (corresponding to a
typical hole mixing $\eta\sim0.2$~\cite{holeMixing}).}%
\label{G_SingleSpin}%
\end{figure}

For $I=1/2$, the motion of the (degree of)\ nuclear polarization
$s(t)\equiv\langle\hat{I}^{z}(t)\rangle/I$ follows from Eq.~(\ref{EOM_GENERAL}%
) as
\begin{equation}
\dot{s}=-\Gamma_{1}s-\Gamma_{p}(s-s_{0}), \label{EOM_S}%
\end{equation}
where $\Gamma_{1}$ is a phenomenological nuclear depolarization rate for other
nuclear spin relaxation mechanisms,
\begin{equation}
\Gamma_{p}=W_{+}+W_{-}=\frac{4\tilde{a}_{h}^{2}}{\gamma_{1}}\left(
\frac{W\gamma_{1}^{2}}{(\gamma_{1}+2W)^{3}}c_{1}+O(\varepsilon^{2})\right)
\label{GAMMAP_GENERAL}%
\end{equation}
is the hole-driven nuclear polarization buildup rate, and%
\begin{equation}
s_{0}=\frac{W_{+}-W_{-}}{W_{+}+W_{-}}=-\frac{\Delta\omega_{N}}{\Delta
^{2}+\gamma_{2}^{2}}\frac{\gamma_{1}}{\gamma_{2}}\frac{c_{0}}{c_{1}%
}+O(\varepsilon^{2}) \label{IZ_GENERAL}%
\end{equation}
is the intrinsic (i.e., when $\Gamma_{1}=0$) steady-state nuclear
polarization, in agreement with Eq.~(\ref{IZ_INTUITIVE}). Here $c_{0}%
\equiv1/2+\gamma_{2}/\gamma_{1}+f+W/\gamma_{1}$ and $c_{1}\equiv1+[\gamma
_{1}/(2\gamma_{2})]f+W/\gamma_{1}$ are non-negative constants, $f\equiv
(\gamma_{2}^{2}-\Delta^{2})/(\gamma_{2}^{2}+\Delta^{2}),$ and $W\equiv
2\pi(\Omega_{R}/2)^{2}L^{(\gamma_{2})}(\Delta)$ is the hole excitation rate.
The steady-state nuclear polarization is $s^{(\mathrm{ss})}=\Gamma_{p}%
s_{0}/(\Gamma_{p}+\Gamma_{1})$. These results agree well with the exact
numerical solutions to the coupled motion (see Fig.~\ref{G_SingleSpin}). For a
general nuclear spin-$I$, the nuclear polarization $s$ still obeys
Eq.~(\ref{EOM_S}) by extending $s_{0}$ to $s_{0}^{(I)}\equiv s_{0}\langle
(\hat{I}^{x})^{2}+(\hat{I}^{y})^{2}\rangle/I$, which reduces to $2(I+1)s_{0}%
/3$ for $|s_{0}^{(I)}|\ll1$.

The second step of the hole-driven collective nuclear stabilization is the
feedback of the nuclear spins on the hole excitation by shifting the energy of
the excited hole state $\left\vert 1\right\rangle $ or the ground state
$\left\vert 0\right\rangle $. For specificity, we consider a negatively
charged QD and identify $\left\vert 1\right\rangle $ with the trion state
(still referred to as \textit{hole}, despite the additional inert electron
spin singlet) and $\left\vert 0\right\rangle $ with the spin-up electron
state. The hole is coupled to the nuclear spins in the QD through the
non-collinear dipolar hyperfine interaction $\hat{\sigma}_{11}\sum_{j}%
\tilde{a}_{h,j}(\hat{I}_{j}^{+}+\hat{I}_{j}^{-})$. The electron is coupled to
the nuclear spins through the diagonal part of the contact hyperfine
interaction $\hat{\sigma}_{00}\sum_{j}a_{e,j}\hat{I}_{j}^{z}/2\equiv
\hat{\sigma}_{00}\hat{h}$, with the off-diagonal terms involving the electron
spin flip being suppressed by the large electron Zeeman splitting. The
feedback of the nuclear spins proceeds by shifting the energy of the electron
state $\left\vert 0\right\rangle $ by $\hat{h}\equiv\sum_{j}a_{e,j}\hat{I}%
_{j}^{z}/2$ (referred to as Overhauser shift hereafter), which consists of the
macroscopic mean-field part $h(t)\equiv\operatorname*{Tr}\hat{P}(t)\hat{h}$
and the fluctuation $\delta\hat{h}(t)\equiv\hat{h}-h(t)$, with $\hat{P}(t)$
being the diagonal part of the reduced density matrix $\hat{\rho}_{N}(t)$ of
the nuclear spins. Then the laser detuning is changed from $\Delta$ to
$\hat{\Delta}\equiv\Delta-\hat{h}\equiv\Delta(t)-\delta\hat{h}(t)$, where
$\Delta(t)\equiv\Delta-h(t)$. This changes the nuclear spin dynamics from
Eq.~(\ref{EOM_GENERAL}) to%
\begin{equation}
\dot{\hat{P}}=-\sum_{j}[\hat{I}_{j}^{-},\hat{I}_{j}^{+}W_{j,+}(\hat{\Delta
})\hat{P}]-\sum_{j}[\hat{I}_{j}^{+},\hat{I}_{j}^{-}W_{j,-}(\hat{\Delta}%
)\hat{P}], \label{EOM_MANY}%
\end{equation}
where $W_{j,\pm}(\hat{\Delta})$ are obtained from $W_{\pm}$ by replacing
$\tilde{a}_{h},\omega_{N},$ and $\Delta$ with $\tilde{a}_{h,j}$, $\omega
_{N,j}$, and $\hat{\Delta}$, respectively. The feedback is manifested in
Eq.~(\ref{EOM_MANY}) as the Overhauser shift $\hat{h}$. As shown below, the
feedback of the macroscopic part $h(t)$ leads to nonlinear, bistable nuclear
spin dynamics, experimentally manifested as hysteretic locking of the
electronic excitation \cite{LattaNaturePhys09}. The feedback of the
fluctuation $\delta\hat{h}(t)$, introduced by qualitative argument
\cite{XuNature09} or stochastic assumption
\cite{OverhauserPR53,reverseOverhauser} in previous treatments, correlates the
dynamics of different nuclear spins and drives the nuclear spins from the
\textquotedblleft noisy\textquotedblright\ thermal equilibrium state into a
\textquotedblleft quiet\textquotedblright\ collective state \cite{XuNature09}.

\begin{figure}[ptb]
\includegraphics[width=\columnwidth]{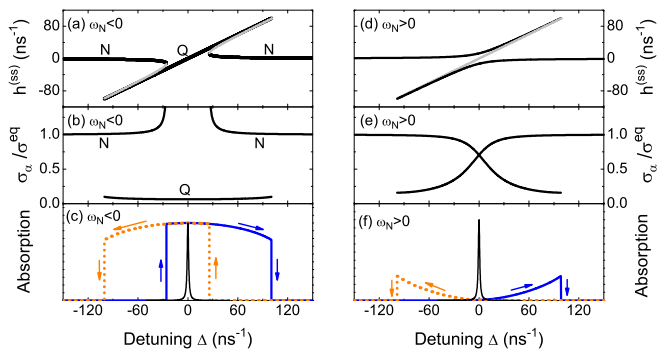}\caption{(color online).
Stable (black lines) and unstable (grey lines) $h^{(\mathrm{ss})}$ [(a), (d)]
and $\sigma_{\alpha}/\sigma^{\mathrm{eq}}$ in stable states [(b),(e)]. (c) and
(f): Optical absorption spectra obtained by sweeping $\Delta$ in different
directions (indicated by the arrows). A typical QD with identical nuclear
spin-9/2's and (unit: $\mathrm{ns}^{-1}$) $Na_{e}=100$, $\gamma_{1}=\gamma
_{2}=\Omega_{R}=1$, and $\omega_{N}=-0.2$ [(a)-(c)] or $0.2$ [(d)-(f)] is
considered. The sharp Lorentzian peaks at $\Delta=0$ in (c) and (f) are
absorption spectra in the absence of the nuclei.}%
\label{G_ManySpin}%
\end{figure}

To keep the exposition simple, we consider identical nuclei $I_{j}=I,$
$\omega_{N,j}=\omega_{N},a_{e,j}=a_{e},\tilde{a}_{h,j}=\tilde{a}_{h}$, so that
$\hat{h}=(Na_{e}I/2)\hat{s}$ ($N$ is the number of QD nuclei) is proportional
to the nuclear polarization operator $\hat{s}\equiv\sum_{j=1}^{N}\hat{I}%
_{j}^{z}/(NI)$. First we consider the feedback of the macroscopic part $h(t)$
by dropping the fluctuation $\delta\hat{h}$ from Eq.~(\ref{EOM_MANY}) to
obtain a mean-field description
\begin{equation}
\dot{h}(t)=-\Gamma_{p}(\Delta(t))[h(t)-h_{0}^{(I)}(\Delta(t))], \label{SCEQ}%
\end{equation}
valid for $I=1/2$ or $|s_{0}^{(I)}|\ll1,$ where $\Gamma_{p}(\Delta(t))$ and
$h_{0}^{(I)}(\Delta(t))$ are obtained from $\Gamma_{p}$ and $h_{0}^{(I)}%
\equiv(Na_{e}I/2)s_{0}^{(I)}$, respectively, by replacing $\Delta$ with
$\Delta(t)=\Delta-h(t)$. Equation~(\ref{SCEQ}) shows that the dynamics of
$h(t)$ is highly nonlinear, since $\Gamma_{p}$ and $s_{0}^{(I)}$ are highly
nonlinear functions of $\Delta$ (see Fig.~\ref{G_SingleSpin}). Consequently
for a given detuning $\Delta,$ the steady-state solution $h^{(\mathrm{ss})}$
may be multi-valued, e.g., two stable solutions (black lines) and one unstable
solution (gray line) in Fig. \ref{G_ManySpin}(a).

For a given detuning $\Delta$, each \textit{stable} solution $h_{\alpha
}^{(\mathrm{ss})}$ corresponds to a macroscopic (mixed) nuclear spin state
with macroscopic polarization $s_{\alpha}^{(\mathrm{ss})}\equiv2h_{\alpha
}^{(\mathrm{ss})}/(Na_{e}I)$. The noise on the electron spin qubit produced by
this state is characterized by the fluctuation of $\hat{s}$ around its
macroscopic value $s_{\alpha}^{(\mathrm{ss})}$. To quantify this fluctuation,
we define the probability distribution function $p(s,t)\equiv
\operatorname*{Tr}\hat{P}(t)\delta(s-\hat{s})$ of $\hat{s}$. Then, instead of
introducing the fluctuation stochastically by assuming that $\hat{s}$
experiences a random walk described by a Fokker-Planck equation (applied to
nuclear spin-1/2's only) \cite{OverhauserPR53,reverseOverhauser}, we treat
this feedback quantum mechanically by directly deriving the equation of motion
(which also assumes the Fokker-Planck form) of $p(s,t)$ from Eq.
(\ref{EOM_MANY}), valid for nuclei with arbitrary spin $I$. The steady-state
solution $p^{(\mathrm{ss})}(s)$ sharply peaks at each macroscopic value
$s_{\alpha}^{(\mathrm{ss})}$. The fluctuation of $\hat{s}$ in the $\alpha$th
macroscopic state is quantified by the standard deviation of the Gaussian peak
at $s_{\alpha}^{(\mathrm{ss})}$:%
\begin{equation}
\sigma_{\alpha}=\sigma^{\text{\textrm{eq}}}\sqrt{\frac{1-(s_{0,\alpha
}^{(\mathrm{ss})})^{2}}{1+dh_{\alpha}^{(\mathrm{ss})}/d\Delta}},
\label{FLUCTUATION}%
\end{equation}
where $s_{0,\alpha}^{(\mathrm{ss})}$ is obtained from $s_{0}$ by replacing
$\Delta$ with $\Delta-h_{\alpha}^{(\mathrm{ss})}$ and $\sigma^{\mathrm{eq}%
}\equiv\left[  (I+1)/(3NI)\right]  ^{1/2}$ is the thermal equilibrium
fluctuation of $\hat{s}$. Equation~(\ref{FLUCTUATION}) is valid for $I=1/2$ or
$|s_{0}^{(I)}|\ll1$ (generalization to arbitrary $s_{0}^{(I)}$ is
straightforward). It unifies the different pictures of nuclear stabilization
(e.g., a qualitative argument of nuclear stabilization by the feedback of
nuclear fluctuation \cite{XuNature09}, nuclear stabilization by a strong
polarization $s^{(\mathrm{ss})}\approx1$~\cite{LossDNP}, and nuclear
stabilization by the competition between polarization and depolarization
processes \cite{OverhauserPR53,reverseOverhauser,VinkNaturePhys09}) by showing
that the degree of stabilization as quantified by the smallness of
$\sigma_{\alpha}/\sigma^{\mathrm{eq}}$ is the product of the individual
contribution $[1-(s_{0,\alpha}^{(\mathrm{ss})})^{2}]^{1/2}$ \cite{LossDNP} and
the collective contribution $(1+dh_{\alpha}^{(\mathrm{ss})}/d\Delta)^{-1/2}$.
The individual contribution comes from the suppression of the fluctuation of
each individual nuclear spin by the steady-state nuclear polarization. The
collective contribution comes from the suppression of the collective
fluctuation of all the nuclear spins by the correlation $(\langle\hat{I}%
_{i}^{z}\hat{I}_{j}^{z}\rangle-\langle\hat{I}_{i}^{z}\rangle\langle\hat{I}%
_{j}^{z}\rangle)/I^{2}\approx-(\sigma^{\mathrm{eq}})^{2}$ between different
nuclear spins, established by the correlated dynamics [Eq. (\ref{EOM_MANY})]
driven by the feedback from the fluctuation $\delta\hat{h}$ \cite{XuNature09}.
The collective contribution typically dominates the nuclear spin
stabilization, e.g., in the Q (i.e., \textquotedblleft quiet\textquotedblright%
) branch in Fig. \ref{G_ManySpin}(b), the nuclear spins are significantly
stabilized collectively, especially near the resonance point $\Delta=0$, where
the slope $dh_{\alpha}^{(\mathrm{ss})}/d\Delta$ (and hence collective
stabilization) is maximal, but the magnitude $|h_{\alpha}^{(\mathrm{ss})}|$
(and hence individual stabilization) is negligible. This explains the
observation by Xu \textit{et al. }in the single pump
experiment~\cite{XuNature09}. For $\omega_{N}>0$ [Fig. \ref{G_ManySpin}(d)],
the nuclear stabilization is maximal at large detunings.

In the Q\ branch of Fig.~\ref{G_ManySpin}(a), $h^{(\mathrm{ss})}\propto\Delta$
always shifts the effective detuning $\Delta^{(\mathrm{ss})}=\Delta
-h^{(\mathrm{ss})}$ towards resonance, while in the N (i.e., \textquotedblleft
noisy\textquotedblright) branch, $h^{(\mathrm{ss})}$ is very small. When
sweeping the laser frequency in different directions, the nuclear spin state
switches between the N\ branch and the Q\ branch, leading to bidirectional
hysteretic locking of the optical absorption peak [Fig. \ref{G_ManySpin}(c)].
This explains the observation by Latta \textit{et al. }%
\cite{LattaNaturePhys09} upon blue exciton excitation when we identify
$\left\vert 0\right\rangle $ as the vacuum and $\left\vert 1\right\rangle $ as
the spin-up exciton. In Fig.~\ref{G_ManySpin}(d), $h^{(\mathrm{ss})}$ always
tends to repel $\Delta^{(\mathrm{ss})}$ away from resonance and the absorption
peak is shifted hysteretically to finite detunings [Fig.~\ref{G_ManySpin}(f)].
Qualitatively similar behaviors are seen in some experimental data under red
exciton excitation~\cite{noteAlexander}.

This research was supported by NSF (PHY 0804114) and U. S. Army Research
Office MURI award W911NF0910406. We thank R. B. Liu, W. Yao, A. H\"{o}gele,
and A. Imamoglu for fruitful discussions. W. Y. thanks M. C. Zhang and Y. Wang
for helpful discussions.


\begin{thebibliography}{99}                                                                                               %


\bibitem {AbragamBook}A. Abragam, \textit{The Principles of Nuclear Magnetism}
(Oxford University Press, New York, 1961).

\bibitem {SpinBath}N. V. Prokof'ev and P. C. E. Stamp, Rep. Prog. Phys.
\textbf{63}, 669 (2000).

\bibitem {Decoherence}M. Schlosshauer, \textit{Decoherence and the
quantum-to-classical transition} (Springer-Verlag, Berlin, 2010).

\bibitem {HoleQC}D. V. Bulaev and D. Loss, Phys. Rev. Lett. \textbf{98},
097202 (2007); A. J. Ramsay \textit{et al.}, \textit{ibid.} \textbf{100},
197401 (2008); B. D. Gerardot \textit{et al.}, Nature \textbf{451}, 441
(2008); D. Brunner \textit{et al.}, Science \textbf{325}, 70 (2009).

\bibitem {HoleEnvironment1}S. Laurent \textit{et al.}, Phys. Rev. Lett.
\textbf{94}, 147401 (2005); D. V. Bulaev and D. Loss, \textit{ibid.}
\textbf{95}, 076805 (2005); D. Heiss \textit{et al.}, Phys. Rev. B
\textbf{76}, 241306(R) (2007).

\bibitem {HoleEnvironment2}J. Fischer \textit{et al.}, Phys. Rev. B\textit{
}\textbf{78}, 155329 (2008); C. Testelin \textit{et al.}, \textit{ibid.
}\textbf{79}, 195440 (2009); B. Eble \textit{et al.}, Phys. Rev. Lett.
\textbf{102}, 146601 (2009); P. Fallahi, S. T. Y\i lmaz, and A. Imamoglu,
\textit{ibid.}, \textbf{105}, 257402 (2010); E. A. Chekhovich \textit{et al.},
\textit{ibid. }\textbf{106}, 027402 (2011).

\bibitem {XuNature09}X. Xu \textit{et al.}, Nature \textbf{459}, 1105 (2009).

\bibitem {LaddPRL10}The sawtooth pattern in electron spin free-induction decay
observed by T. D. Ladd \textit{et al. }[Phys. Rev. Lett. \textbf{105}, 107401
(2010)] is attributed to a similar process.

\bibitem {ElectronQC}R. Hanson \textit{et al.}, Rev. Mod. Phys. \textbf{79},
1217 (2007); T. D. Ladd, \textit{et al.}, Nature \textbf{464}, 45 (2010); R.
B. Liu, W. Yao, and L. J. Sham, Adv. Phys. \textbf{59}, 703 (2010).

\bibitem {DecoherenceNuclei}R. de Sousa, and S. Das Sarma, Phys. Rev. B,
\textbf{68}, 115322 (2003); R. B. Liu, W. Yao, and L. J. Sham, New J. Phys.
\textbf{9}, 226 (2007); J. Fischer \textit{et al.}, Solid State Commun.
\textbf{149}, 1443 (2009).

\bibitem {DD}For a review, see W. Yang, Z. Y. Wang, and R. B. Liu, Front.
Phys. \textbf{6}, 2 (2011).

\bibitem {LossDNP}W. A. Coish and D. Loss, Phys. Rev. B, \textbf{70}, 195340 (2004).

\bibitem {OverhauserPR53}A. W. Overhauser, Phys. Rev. \textbf{92}, 411 (1953);
J. Danon and Y. V. Nazarov, Phys. Rev. Lett. \textbf{100}, 056603 (2008).

\bibitem {reverseOverhauser}E. A. Laird \textit{et al.}, Phys. Rev. Lett.
\textbf{99}, 246601 (2007); M. S. Rudner and L. S. Levitov, \textit{ibid.}
\textbf{99}, 246602 (2007); J. Danon \textit{et al.}, \textit{ibid.}
\textbf{103}, 046601 (2009).

\bibitem {GreilichScience07}A. Greilich \textit{et al.}, Science \textbf{317},
1896 (2007).

\bibitem {ReillyScience08}D. J. Reilly \textit{et al.}, Science \textbf{321},
817 (2008).

\bibitem {LattaNaturePhys09}C. Latta \textit{et al.}, Nature Phys. \textbf{5},
758 (2009).

\bibitem {VinkNaturePhys09}I. T. Vink \textit{et al.}, Nature Phys.
\textbf{5}, 764 (2009).

\bibitem {ImamogluNarrowing}Issler \textit{et al.}[arXiv:cond-mat/1008.3507v1]
propose that electron-driven nuclear spin diffusion stabilizes the nuclear
spins in the two-pump configuration of Ref. [\onlinecite{XuNature09}].

\bibitem {holeMixing}B. Eble \textit{et al.}, Phys. Rev. Lett. \textbf{102},
146601 (2009). Dreiser \textit{et al}. [Phys. Rev. B \textbf{77}, 075317
(2008)] found a small optically active mixing $\eta_{\mathrm{active}}%
\approx0.02$.

\bibitem {noteAlexander}A. H\"{o}gele, private communication.
\end{thebibliography}
\end{document}